\documentclass[twoside,leqno]{article}

\usepackage[letterpaper]{geometry}

\usepackage{siamproceedings}
\usepackage[numbers]{natbib}
\usepackage[T1]{fontenc}
\usepackage{amsfonts}
\usepackage{graphicx}
\usepackage{epstopdf}
\usepackage{enumitem}
\usepackage{algorithmic}
\usepackage{graphicx}
\usepackage{caption}
\usepackage{subcaption}
\ifpdf
  \DeclareGraphicsExtensions{.eps,.pdf,.png,.jpg}
\else
  \DeclareGraphicsExtensions{.eps}
\fi

\newsiamremark{remark}{Remark}
\newsiamremark{hypothesis}{Hypothesis}
\crefname{hypothesis}{Hypothesis}{Hypotheses}
\newsiamthm{claim}{Claim}

\usepackage{amsopn}

\begin{document}

\newcommand\relatedversion{}
\renewcommand\relatedversion{\thanks{The full version of the paper can be accessed at \protect\url{https://arxiv.org/abs/0000.00000}}} 

\title{\Large 
GRU-Based Learning for the Identification of Congestion Protocols in TCP Traffic
}
    \author{Paul Bergeron\thanks{School of Computer Science and Mathematics, Marist University, Poughkeepsie, NY, USA (\email{Paul.Bergeron1@marist.edu}).}
    \and Sandhya Aneja\thanks{School of Computer Science and Mathematics, Marist University, Poughkeepsie, NY, USA (\email{sandhya.aneja@marist.edu}.)}
}

\date{}

\maketitle


\fancyfoot[R]{\scriptsize{Copyright \textcopyright\ 2025 by SIAM\\
Unauthorized reproduction of this article is prohibited}}





\begin{abstract} This paper presents the identification of congestion control protocols TCP Reno, TCP Cubic, TCP Vegas, and BBR on the Marist University campus, with an accuracy of 97.04\% using a GRU-based learning model. We used a faster neural network architecture on a more complex and competitive network in comparison to existing work and achieved comparably high accuracy.

\end{abstract}

\newcommand{\accuracy}{97.04\%}
\newcommand{\days}{15}
\newcommand{\features}{size, maximum window size, throughput, smoothed throughput and RTT}
\newcommand{\RTTs}{47}
\newcommand{\TCPCCFirst}{TCP Congestion Control algorithm (CC)}
\newcommand{\TCPCC}{TCP CC}

\section{Introduction}
Existing research has shown the need to identify \TCPCCFirst \cite{6594906}. However, with the growing Internet, as different organizations are adopting different congestion control algorithms, there is a need to reevaluate how to identify the \TCPCC. The \TCPCC~ identification problem is the identification of the congestion control algorithm used by a web server while sending the requested resource, such as a file, a webpage, or a record from the database. In many cases, sending the requested resource by a server involves fetching the resource from other servers by cascading requests again using \TCPCC~ by different servers. Identification of the \TCPCC~ can help network administrators to measure the throughput, delay, loss, and jitter for network performance since the congestion control algorithm largely determines these.

The \TCPCC~ identification is valuable not only in measuring network performance, but also helps in other network applications, such as device fingerprinting for device authentication, browser fingerprinting for web services, and web server fingerprinting for cybersecurity. These applications utilize features from network traffic between a device and the web server for fingerprinting purposes. In this paper, we suggest the \TCPCC~ itself as a feature for the network traffic analysis-based applications. 

We found four algorithms in Ubuntu 20.04 Linux kernel, namely TCP Vegas, TCP Reno, TCP Cubic \citep{rhee2018cubic}, and BBRv1(Bottleneck Bandwidth and Round Trip Time)\citep{cardwell2016bbr}, to set up a web server. We chose an RNN-based Gated Recurrent Unit (GRU) architecture by looking into the communications by setting congestion control protocols. We collected the data over \days~ days on the Marist University Campus over three times in a day and achieved \accuracy~ accuracy in identifying the congestion control algorithm. Our contributions in this work are as follows:
\begin{itemize}
  \item We achieve \accuracy~ accuracy in identifying congestion control algorithms using an RNN-based GRU model.

\item We identified Congestion control as a representative feature representing \features~ for network analysis. 

\item We identified that the bottleneck link at Marist Campus is 1Gbps at the Hancock building, and the maximum throughput can be achieved by BBRv1, while the minimum round-trip time can be achieved by TCP Cubic.
\end{itemize}

\begin{figure}[htbp]
    \centering
    \setlength\fboxsep{1pt} 
    \setlength\fboxrule{0.5pt}

    \begin{subfigure}[b]{0.23\textwidth}
        \centering
        \fbox{\includegraphics[width=\linewidth]{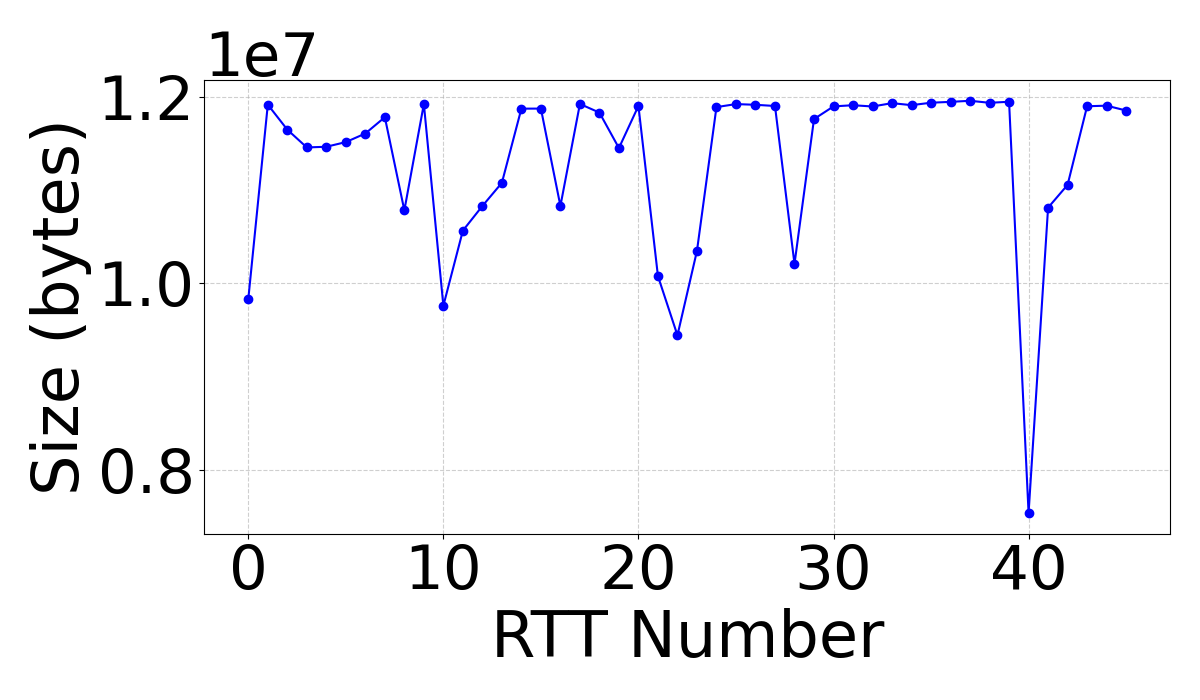}}
        \caption{Size - BBR}
    \end{subfigure}
    \hfill
    \begin{subfigure}[b]{0.23\textwidth}
        \centering
        \fbox{\includegraphics[width=\linewidth]{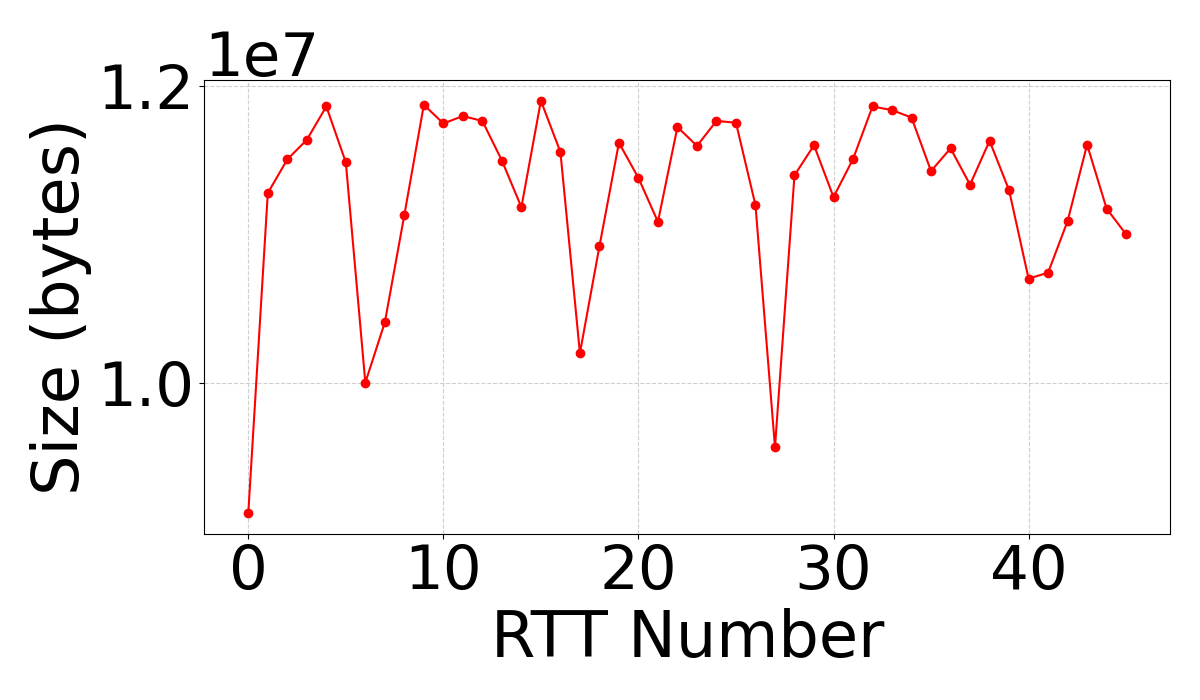}}
        \caption{Size - CUBIC}
    \end{subfigure}
    \hfill
    \begin{subfigure}[b]{0.23\textwidth}
        \centering
        \fbox{\includegraphics[width=\linewidth]{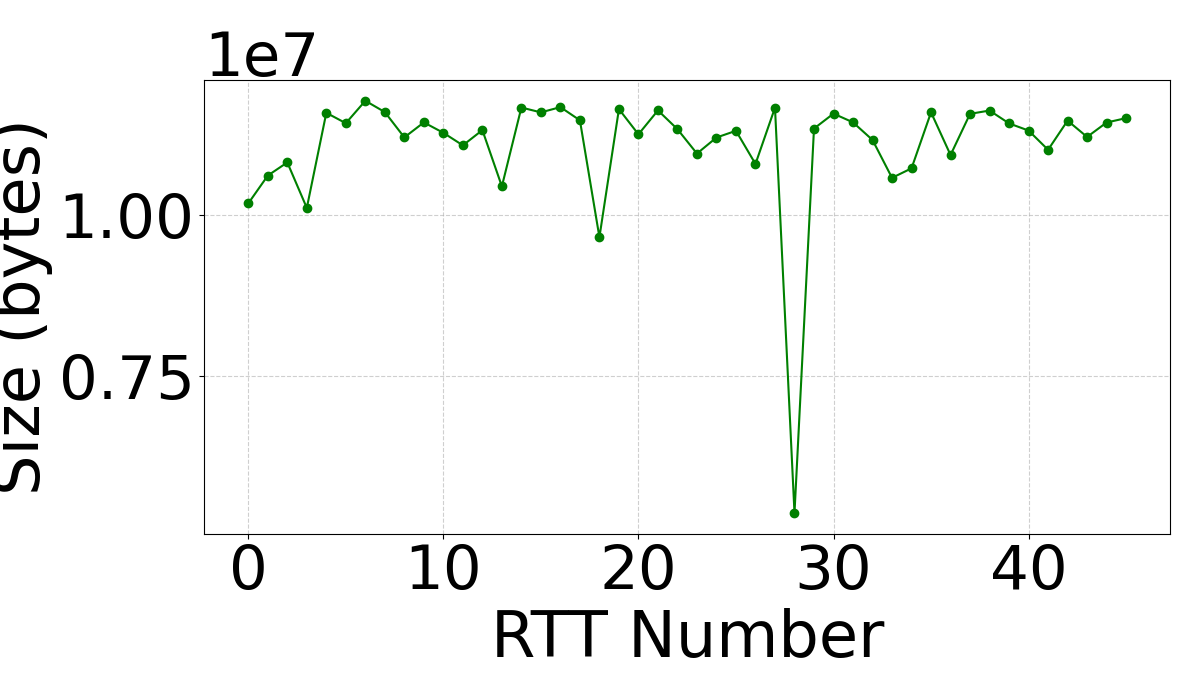}}
        \caption{Size - RENO}
    \end{subfigure}
    \hfill
    \begin{subfigure}[b]{0.23\textwidth}
        \centering
        \fbox{\includegraphics[width=\linewidth]{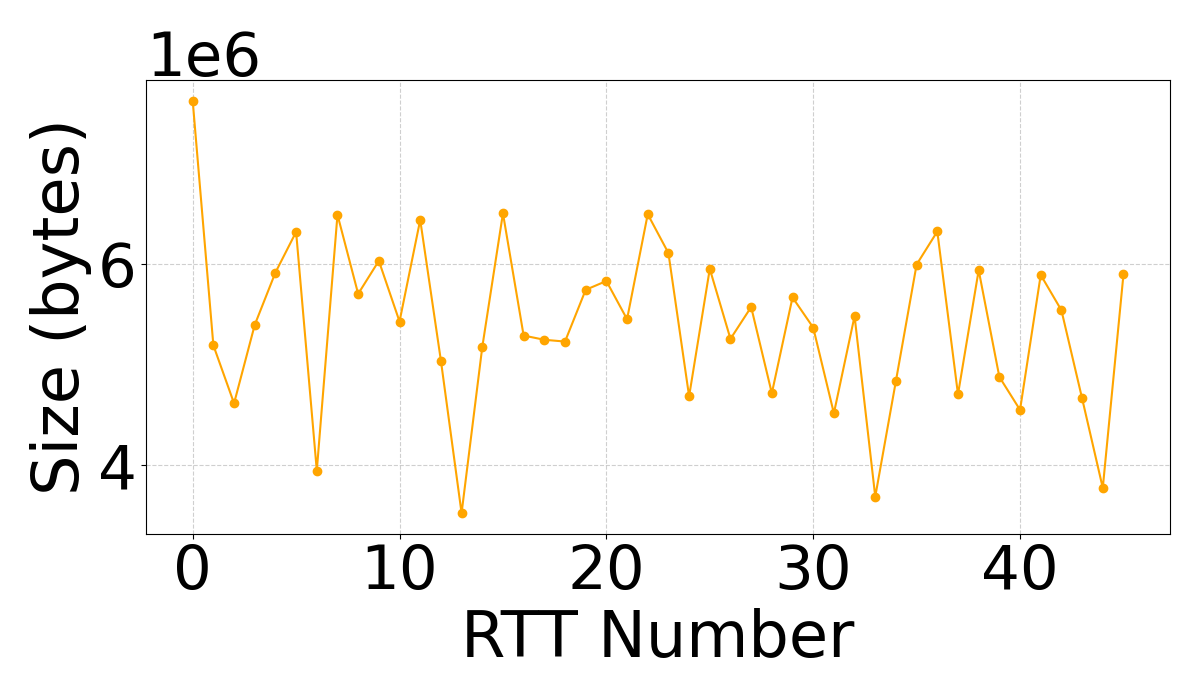}}
        \caption{Size - VEGAS}
    \end{subfigure}

    \vspace{0.3cm}

    \begin{subfigure}[b]{0.23\textwidth}
        \centering
        \fbox{\includegraphics[width=\linewidth]{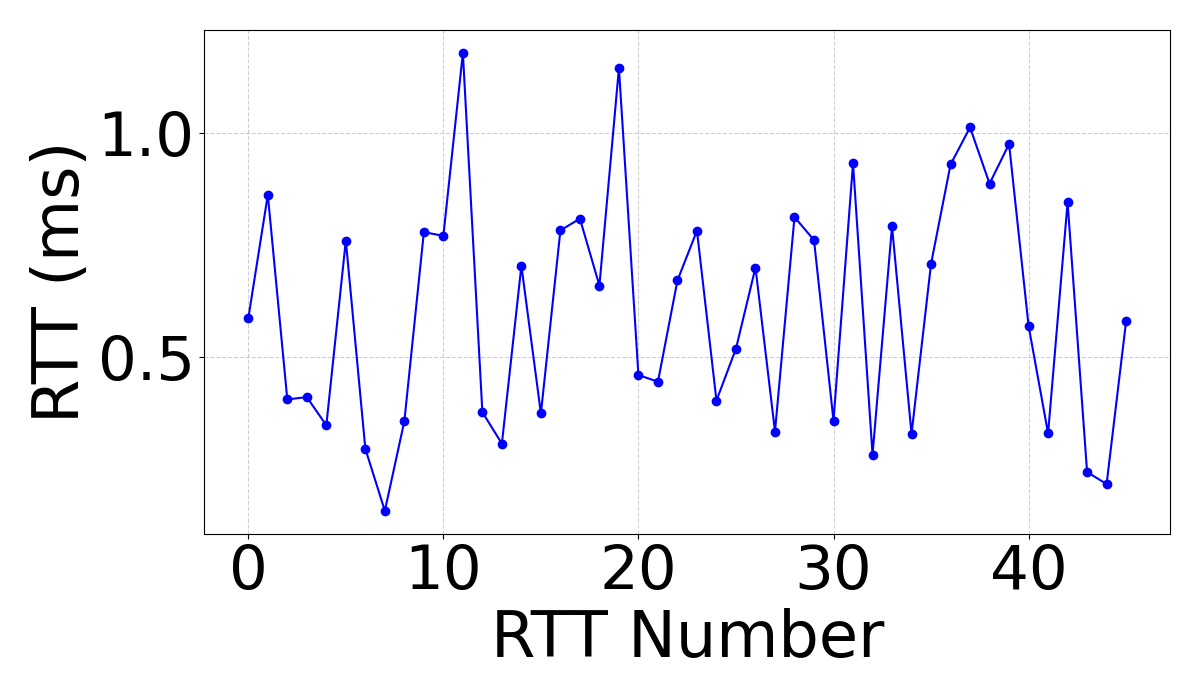}}
        \caption{RTT - BBR}
    \end{subfigure}
    \hfill
    \begin{subfigure}[b]{0.23\textwidth}
        \centering
        \fbox{\includegraphics[width=\linewidth]{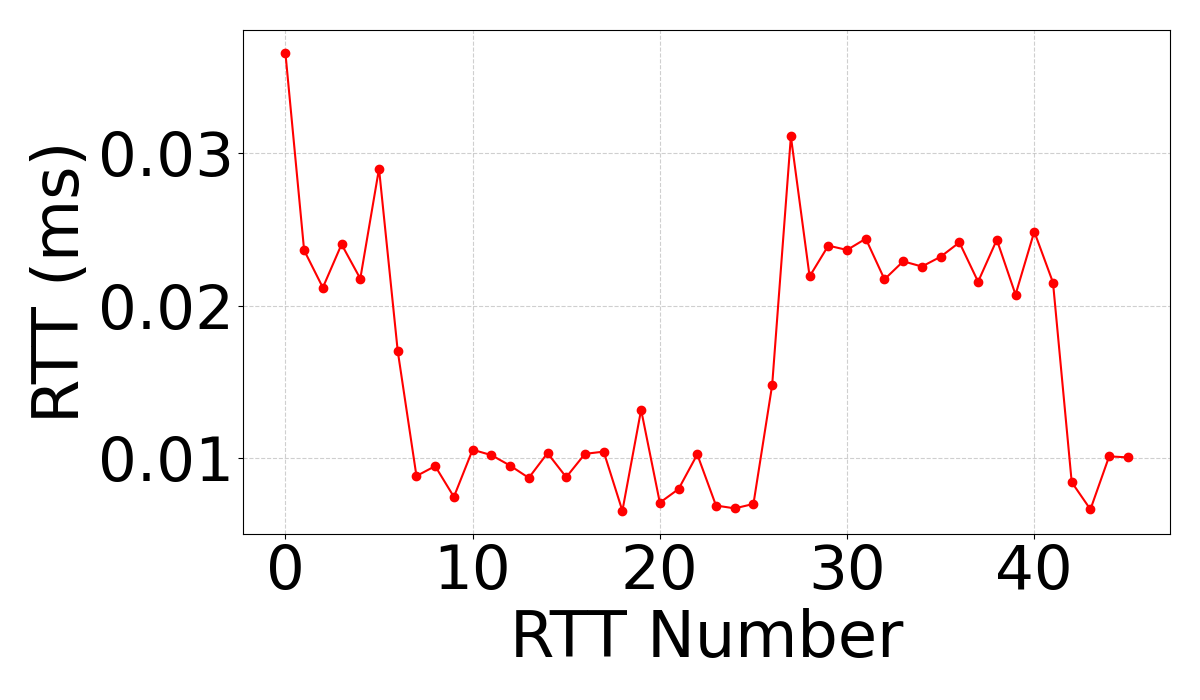}}
        \caption{RTT - CUBIC}
    \end{subfigure}
    \hfill
    \begin{subfigure}[b]{0.23\textwidth}
        \centering
        \fbox{\includegraphics[width=\linewidth]{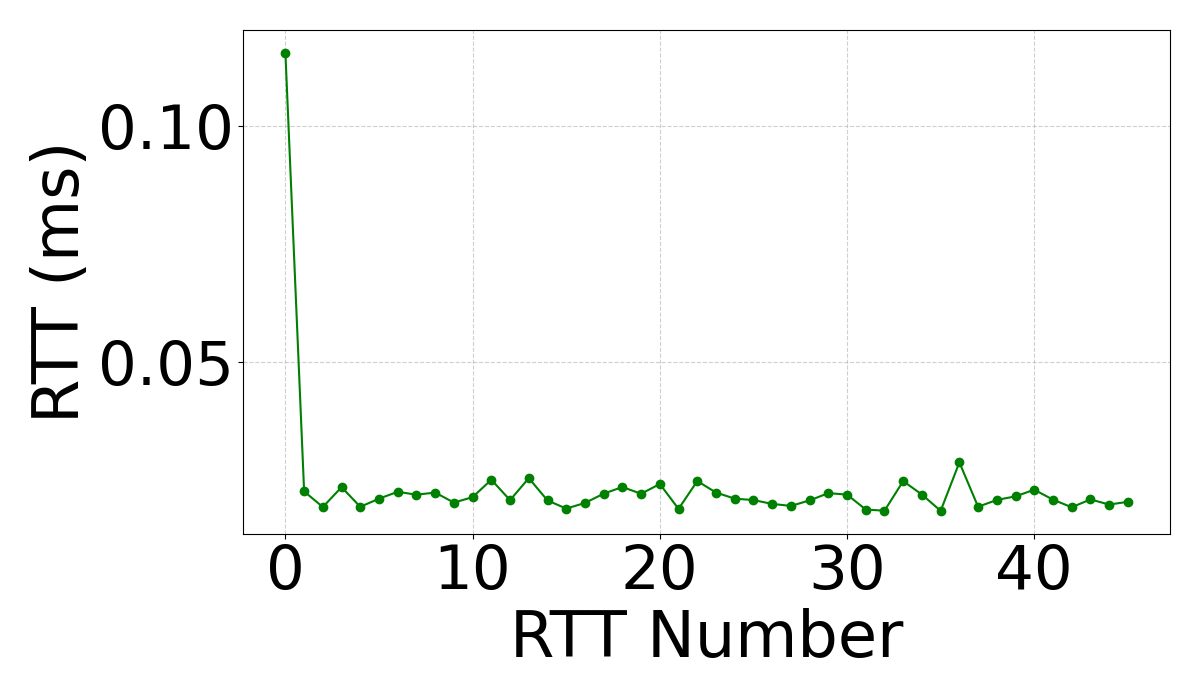}}
        \caption{RTT - RENO}
    \end{subfigure}
    \hfill
    \begin{subfigure}[b]{0.23\textwidth}
        \centering
        \fbox{\includegraphics[width=\linewidth]{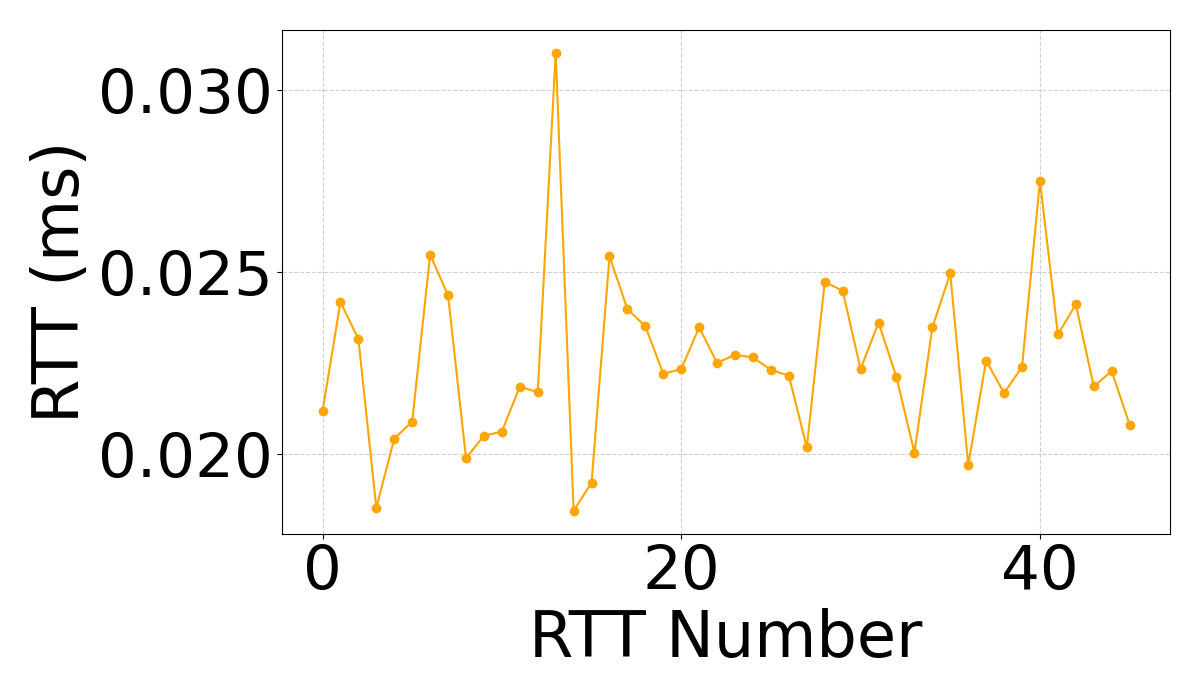}}
        \caption{RTT - VEGAS}
    \end{subfigure}

    \caption{Size and RTT Variation for BBR, CUBIC, RENO, and VEGAS}
    \label{fig:all_8_plots}
\end{figure}

\section{Motivation}
\TCPCC~ is governed by the algorithms outlined in RFC 5681, which formalizes the use of slow start, congestion avoidance, fast retransmit, and fast recovery to adjust the sending rate based on packet loss signals \cite{allman2009tcp}.
TCP Reno reduces the congestion window by half upon detecting a packet loss and updates the threshold by window size at that time. Reno increases its window linearly with each acknowledgment (ACK) received, a behavior known as additive increase. TCP Cubic improves Reno by using a cubic function of time since the last window reduction to adjust its window size. Cubic uses more aggressive growth when the network is underutilized and smoother convergence near the saturation point. TCP Vegas estimates the expected throughput based on the minimum RTT and compares it with the actual throughput. Vegas uses two thresholds, $\alpha$ and $\beta$: if the difference between expected and actual throughput exceeds $\beta$, it interprets this as a sign of congestion and reduces the window. If the difference is less than $\alpha$, it increases the window, assuming the network is underutilized. BBR explores the network's bandwidth and delay characteristics by exponentially increasing the window during a startup phase. Upon reaching a bandwidth plateau, BBR transitions to a drain phase to reduce in-flight data, followed by a cycle of phases including probeBW and probeRTT. BBR aims to maintain the congestion window around three times the bandwidth-delay product to maximize throughput while controlling latency.

These protocols differ significantly in how they increase and decrease the congestion window across RTTs. However, for a given protocol, the window adjustment behavior remains consistent across different communication runs. This consistency provides an insight into applying machine learning to effectively identify and classify protocol behavior based on observed patterns in network traffic. 

Figure~\ref{fig:all_8_plots} supports our earlier observations. Subfigures (a), (b), (c), and (d) display the variation in the number of bytes sent (size) per RTT for each of the four protocols. Among them, Figure~\ref{fig:all_8_plots}a (BBR) shows the highest throughput, followed by Cubic, Reno, and Vegas. Subfigures (e), (f), (g), and (h) illustrate the RTT variation over the same \RTTs~RTTs. These plots reveal that Cubic and Reno maintain consistently lower RTTs compared to BBR and Vegas. Notably, BBR exhibits significantly higher RTT values, likely due to its aggressive probing behavior and larger in-flight data volume. These experimental results further confirm that each congestion control algorithm follows a distinct mechanism for adjusting the congestion window over time, reinforcing the feasibility of protocol differentiation based on observed traffic patterns.

\section{Problem Statement}
We address the problem of distinguishing among \( m \) communication flows, denoted \( c_1, c_2, \ldots, c_m \), where each flow operates under one of \( k \) possible congestion control protocols \( p_1, p_2, \ldots, p_k \).

Consider the \( i^{\text{th}} \) communication flow \( c_i \), which uses protocol \( p_i \) and completes \( n \) round trips, represented by \( ct_1, ct_2, \ldots, ct_n \). We hypothesize that if another communication flow \( c_j \) also uses the same protocol \( p_i \), then the sequences of round-trip times for \( c_i \) and \( c_j \) will exhibit similar temporal patterns. Furthermore, if \( b_x \) bytes are transmitted during round trip \( ct_x \) of \( c_i \), as determined by \( p_i \), then the number of bytes transmitted during a corresponding round trip \( ct_y \) of \( c_j \) will also reflect the same protocol-governed behavior.

To learn these patterns, we employ a GRU model with an attention mechanism that captures the temporal dependencies inherent in the behavior of congestion control protocols. GRUs are well-suited for such tasks, as they are designed to capture patterns over time and retain information from earlier inputs that influence later outputs. For example, the characteristic sawtooth pattern of TCP Reno, or the gradual bandwidth ramp-up in BBR, cannot be identified from a single snapshot—they require modeling of temporal evolution. Thus model is required to train to learn a set of input-output pairs \( (x, y) \), where \( x \) is a sequence comprising \features, and \( y \) is the corresponding congestion control protocol \( p_i \). 
Additionally, GRUs are computationally efficient, less prone to overfitting on small to mid-sized datasets, and maintain competitive performance compared to LSTMs. They also integrate well with attention mechanisms, further enhancing temporal pattern recognition.

\section{Literature Review}
 \citet{pahdye2001inferring} presented a tool, TBIT—TCP Behavior Inference Tool which uses heuristic rules-based method to identify congestion control protocol. TBIT checks how a web server's TCP behaves—whether it follows rules, uses an updated congestion control method. By simulating packet drops and observing how the server reacts, TBIT can distinguish which version of TCP is in use.

 \citet{6594906} proposed CAAI—Congestion Avoidance Algorithm Identification—a tool to actively probe remote web servers with the goal of identifying their TCP congestion avoidance algorithms. By emulating network environments and extracting TCP features, CAAI made a large-scale measurement of about 30,000 servers, revealing a shift from Reno to heterogeneous congestion control algorithms such as Cubic. However, active probing is limited applicably in passive or encrypted contexts.

 \citet{sander2019deepcci} presented DeepCCI—Deep Learning-based Passive Congestion Control Identification. DeePCCI uses the arrival time of packets in a flow, grouped into equal time intervals (like buckets), forming a histogram that shows how many packets arrived in each interval. They use a CNN+LSTM architecture combining 1D CNN layers for feature extraction and LSTM layers for capturing temporal patterns in packet arrival histograms. They tested 2-50 Mbps bandwith network for 0-50 ms delay and reported accuracy of 99\%.  
 
Our work extends prior work by applying GRU with attention mechanism, and analyzing more variables on a 1 Gbps bandwith campus network with 0-0.09 ms delay to study identification accuracy and robustness in realistic, heterogeneous network conditions.

\section{Experimental Setup} 
We employ a client–server architecture, with the server hosted on a virtual machine in our enterprise computing laboratory at the Hancock building, Marist University, connected to the network through a 1 Gbps bottleneck link. The client computer runs a script that connects to server using ssh with admin permission to change the congestion control protocol. The script is configured to change the protocols TCP Vegas, TCP Reno, TCP Cubic, and BBRv1 at the server. After each change of protocol, the client downloaded 500 MB data file from the server using http. This download of the data is captured by Wireshark in a pcap file. The client script repeated this pattern using crontab autonomously every day at 6 am, 12 noon, and 6 pm for \days~ days. The pcap files were labeled by the time of capture and the protocol used. After capturing the data in the pcap files, we used tshark in Python to extract features such as \features. 
\begin{figure}[htbp]
    \centering
   \includegraphics[width=0.6\textwidth,height=0.2\textheight]{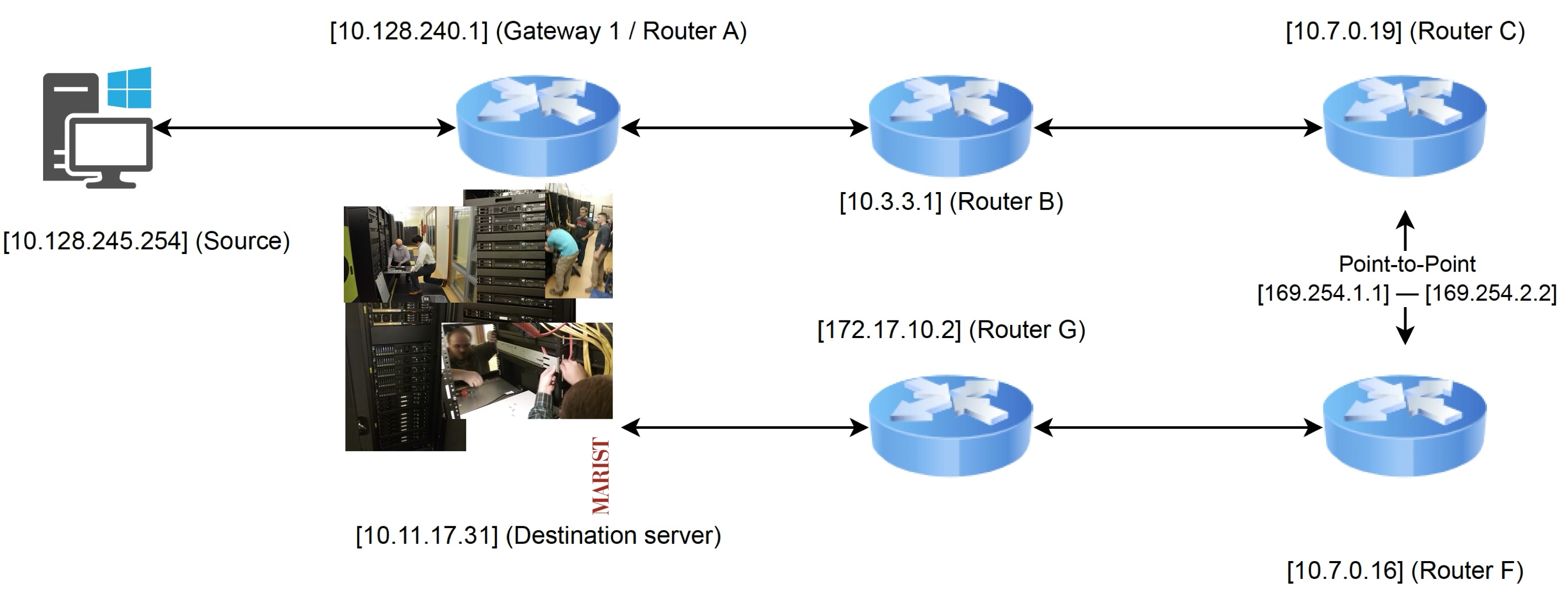}
    \caption{Testbed.}
    \label{fig:testbed}
\end{figure}

\subsection{Features}
The dataset captured detailed TCP flow characteristics recorded at regular intervals of 100 milliseconds during active communication sessions. Each entry includes the timestamp (time), volume of data transmitted (size), maximum allowed congestion window size (Max\_Winc), instantaneous throughput in megabits per second (Mbps), a smoothed throughput value (Smoothed) for noise reduction, and the measured round-trip time (rtt\_ms). The smoothed column was calculated by first computing the raw throughput per interval, then with the moving average using python library pandas ".rolling().mean()."
The protocol is inferred from the file pcap name, where it was recorded previously. The time was not included as features for machine learning. 

\subsection{GRU Model}
We employed GRU-based neural network with three layers and a hidden size of 512. The model is bidirectional, enabling it to capture both forward and backward temporal dependencies. A dropout probability of 0.4 was applied between layers to mitigate overfitting. 
The model was trained using cross-entropy loss function. We employed the Adam optimizer with a learning rate of 0.000075. To improve convergence, a learning rate scheduler was applied using \texttt{ReduceLROnPlateau}, configured with a reduction factor of 0.5 and patience of 5 epochs. Training was conducted over 30 epochs with a batch size of 8. Model performance was evaluated using two metrics: classification accuracy (reported in percentage) and cross-entropy loss.  The balanced data samples were then distributed among the training, validation, and final tests in the ratio of 0.7:0.1:0.2, respectively. We used sequence length of 60 time steps.

\section{Results}
\begin{figure}[htbp]
    \centering
    \begin{minipage}[b]{0.45\textwidth}
        \centering
        \begin{tabular}{ |c|p{1cm}|p{1cm}|p{1cm}|p{1cm}| } 
         \hline
         Protocol: & TCP Vegas & TCP Reno & TCP Cubic & BBR \\ 
         Samples: & 3221 & 1802 & 1777 & 1629 \\  
         \% of total & 38.6\% & 21.6\% & 21.3\% & 19.5\% \\
         \hline
         \multicolumn{5}{|c|}{%
           \parbox{0.9\linewidth}{\centering
             \textbf{Overall accuracy: 97.4\%}\\
             Bandwidth: 1 Gbps \textbar{} Delay: 0.09--0.10 ms
           }%
         }\\
         \hline
        \end{tabular}
        \captionof{table}{Distribution of samples by protocol.}
        \label{tab:data_split}
    \end{minipage}
    \hfill
    \begin{minipage}[b]{0.5\textwidth}
        \centering
        \includegraphics[width=\textwidth]{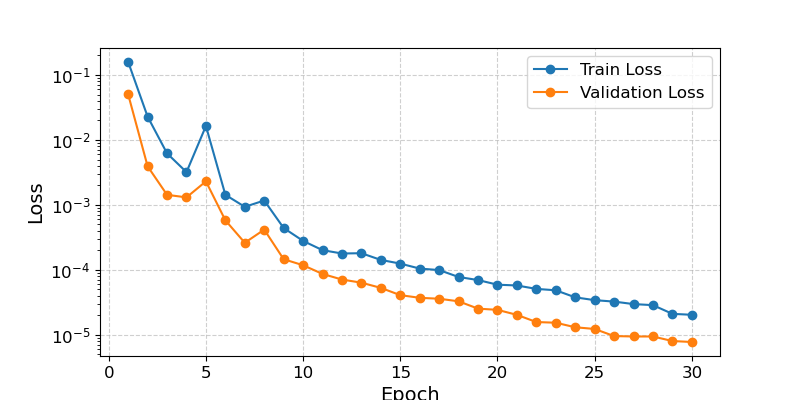}
        \caption{Training and validation loss (log scale).}
        \label{fig:loss_plot}
    \end{minipage}

\end{figure}
The table \ref{tab:data_split} shows the samples per protocol with a breakdown by percentage. The dataset, although collected over time, shows a notable imbalance: TCP Vegas accounts for 38.6\% of the samples, while the other protocols range between 19.5\% and 21.6\%. This imbalance arises from differences in protocol behavior, as the protocol achieving the highest throughput completes its transmissions first. To address this, we standardized the dataset by retaining the same number of rows for all protocols as the one that finished earliest. The paper employs an attention mechanism to handle shifts in context that occur when protocols change. Attention is particularly effective in capturing such contextual transitions within the data.

Figure \ref{fig:loss_plot} shows the training and validation loss over epochs, which demonstrates that the training and validation loss gradually decreased and the model converged. After the model converged, we got the test accuracy of 97.04\%, which is over a potentially more complex and competitive network environment. 

\section{Conclusion} We considered the communication metadata in our experimental setup, such as the data size per second (in bytes), window size, and RTT. Consequently, the results remain valid even for encrypted traffic. It is important to note that different environments exhibit distinct characteristics. For instance, data centers and Ethernet networks impose stringent delay requirements, while wireless networks such as WiFi and cellular systems introduce varying throughput and RTT behaviors. These variations can influence the achievable accuracy of the results.

\bibliographystyle{plainnat}
\bibliography{references}
\end{document}